\documentclass[11pt,a4paper]{article}

\usepackage{amsmath}
\usepackage{amstext}
\usepackage{amssymb}
\usepackage{science}
\usepackage{bm}
\usepackage{color}
\RequirePackage[numbers,sort&compress]{natbib}

\def\ihalf{\textstyle{\frac{i}{2}}}

\def\onehalf{\textstyle{\frac{1}{2}}}

\def\D{{\mathcal D}{}}
{}

\def\Abol{{\stackrel{~\circ}{A}}{}}

\def\Rbol{{\stackrel{\circ}{R}}{}}
\def\Lbol{{\stackrel{\circ}{\mathcal L}}{}}
\def\Tbol{{\stackrel{\circ}{T}}{}}

\def\Gammaw{{\stackrel{\bullet}{\Gamma}}{}}
\def\gammaw{{\stackrel{\bullet}{\gamma}}{\hspace{0.8pt}}}

\def\Rw{{\stackrel{\bullet}{R}}{}}

\def\jw{{\stackrel{\;\bullet}{J}}{}}

\def\tw{{\stackrel{\bullet}{t}}{}}

\def\Lw{{\stackrel{~\bullet}{\mathcal L}}{}}
\def\Tw{{\stackrel{\bullet}{T}}{}}

\def\Kw{{\stackrel{\bullet}{K}}{}}
\def\Aw{{\stackrel{~\bullet}{A}}{}}

\def\nablaw{{\stackrel{\bullet}{\nabla}}{}}
\def\Dw{{\stackrel{\bullet}{\mathcal D}}{}}
\def\Sw{{\stackrel{\bullet}{\mathcal S}}{}}

\def\sw{{\stackrel{\bullet}{S}}{}}

\def\Sw{{\stackrel{\bullet}{\mathcal S}}{}}

\def\be{\begin{equation}}
\def\ee{\end{equation}}
\def\ba{\begin{eqnarray}}
\def\ea{\end{eqnarray}}

\usepackage[colorlinks=true, pdfstartview=FitV, linkcolor=blue, 
            citecolor=blue, urlcolor=blue]{hyperref}

\begin{document}

\begin{center}
{\Large \bf Teleparallelism: A New Insight Into Gravity}\footnote{Chapter in \href{http://www.springer.com/physics/theoretical,+mathematical+&+computational+physics/book/978-3-642-41991-1}{\em Springer Handbook of Spacetime}, edited by A. Ashtekar and V. Petkov (Springer, Berlin, 2014).}
\vskip 0.5cm
{\bf J. G. Pereira}
\vskip 0.03cm 
{\it Instituto de F\'{\i}sica Te\'orica, Universidade Estadual Paulista \\
Caixa Postal 70532-2, 01156-970 S\~ao Paulo, Brazil}
\end{center}
\vskip 0.5cm
\begin{quote}
{\footnotesize Teleparallel gravity, a gauge theory for the translation group, turns up as fully equivalent to Einstein's general relativity. In spite of this equivalence, it provides a whole new insight into gravitation. It breaks several paradigms related to the geometric approach of general relativity, and introduces new concepts in the description of the gravitational interaction. The purpose of this chapter is to explore some of these concepts, as well as discuss possible consequences for gravitation, mainly those that could be relevant for the quantization of the gravitational field.

}
\end{quote}

\null \vskip 0.8cm
\tableofcontents

\newpage
\section{Preliminaries}

Despite being equivalent to general relativity, teleparallel gravity is, conceptually speaking, a completely different theory. For example, the gravitational field in this theory is represented by torsion, not by curvature. Furthermore, in general relativity curvature is used to {\it geometrize} the gravitational interaction: geometry replaces the concept of gravitational force, and the trajectories are determined by geodesics --- trajectories that follow the curvature of spacetime. Teleparallel gravity, on the other hand, attributes gravitation to torsion, which acts as a {\it force}, not geometry. In teleparallel gravity, therefore, trajectories are not described by geodesics, but by force equations \cite{AndPer97}.

The reason for gravitation to present two equivalent descriptions is related to its most peculiar property: {\em universality}. Like the other fundamental interactions of nature, gravitation can be described in terms of a gauge theory. This is just teleparallel gravity, a gauge theory for the translation group. Universality of free fall, on the other hand, allows a second, geometric description, based on the equivalence principle, just general relativity. As the unique universal interaction, it is the only one to allow a geometric interpretation, and hence two alternative descriptions. From this point of view, curvature and torsion are simply alternative ways of representing the very same gravitational field, accounting for the same degrees of freedom of gravity (There are models in which curvature and torsion are related to different degrees of freedom of gravity. In these models, known as Einstein-Cartan-Sciama-Kibble theories, in addition to energy and momentum, also intrinsic spin appears as source of gravitation. The main references on these theories can be traced back from Ref.~\cite{HelhMilu}.) 

The notion of teleparallel structure --- also known as absolute or distant parallelism, characterized by a particular Lorentz connection that parallel-transports everywhere the tetrad field (See Section~\ref{SpinConne} for a remark about the notion of absolute parallelism condition and local Lorentz transformations) --- was used by Einstein in his unsuccessful attempt to construct a unified field theory of electromagnetism and gravitation \cite{UniEin}. The birth of teleparallel gravity as a gravitational theory, however, took place in the late fifties and early sixties with the works by M{\o}ller \cite{moller}. Since then many contributions from different authors have been incorporated into the theory, giving rise to what is known today as the teleparallel equivalent of general relativity, or just teleparallel gravity \cite{livro2}. The purpose of this chapter is to review the fundamentals of this theory, as well as to explore some of the new insights it provides into gravitation, in particular those that could eventually be relevant for the development of a quantum theory for gravitation. 

\section{Basic Concepts}

\subsection{Linear Frames and Tetrads}
\label{sec:framestetrads} \index{Tetrad field}

Spacetime is the arena on which the four presently known fundamental interactions manifest themselves. Electromagnetic, weak and strong interactions are described by gauge theories involving transformations taking place in {\em internal} spaces, by themselves unrelated to spacetime. The basic setting of gauge theories are the principal bundles, in which a copy of the gauge group is attached to each point of spacetime --- the base space of the bundle. Gravitation, on the other hand, is deeply linked to the very structure of spacetime. The geometrical setting of gravitation is the tangent bundle,\index{Tangent bundle} a natural construction always present in any differentiable manifold: at each point of spacetime there is a tangent space attached to it --- the fiber of the bundle --- which is seen as a vector space. We are going to use the Greek alphabet $(\mu, \nu, \rho, \dots = 0,1,2,3)$ to denote indices related to spacetime, and the first letters of the Latin alphabet $(a,b,c, \dots = 0,1,2,3)$ to denote indices related to the tangent space, a Minkowski spacetime whose Lorentz metric is assumed to have the form
\be
\eta_{ab} = \mathrm{diag}(+1,-1,-1,-1).
\label{eq:etaofMinko}
\ee

A general spacetime is a 4-dimensional differential manifold, denoted ${\mathbb R}^{3,1}$, whose tangent space is, at each point, a Minkowski spacetime. Spacetime coordinates will be denoted by $\{x^\mu\}$, whereas tangent space coordinates will be denoted by $\{x^a\}$. Such coordinate systems determine, on their domains of definition, local bases for vector fields, formed by the sets of gradients
\be
\{\partial_\mu\} \equiv \{ {\partial}/{\partial x^\mu} \} \quad \mbox{and} \quad
\{\partial_a\} \equiv \{ {\partial}/{\partial x^a} \},
\ee
as well as bases $\{dx^\mu\}$ and $\{dx^a\}$ for covector fields, or differentials. These bases are dual, in the sense that
\be
dx^\mu ({\partial_\nu}) = \delta^\mu_\nu \quad \mbox{and} \quad
dx^a ({\partial_b}) = \delta^a_b.
\ee
On the respective domains of definition, any vector or covector can be expressed in terms of these {\em coordinate bases}, a name that stems from their relationship to a coordinate system.

\subsubsection{Trivial Frames}
\label{sec:frames}  \index{Tetrad field!trivial}

Trivial frames, or trivial tetrads \cite{livro}, will be denoted by 
\be
\{e_{a}\} \quad \mbox{and} \quad \{e^{a}\}.
\ee
The above mentioned coordinate bases 
\be
\{{e_a}\} = \{{\partial_a}\} \quad \mbox{and} \quad
\{{e^a}\} = \{{d x^a}\}
\ee
are very particular cases. Any other set of four linearly independent fields $\{e_{a}\}$ will form another basis, and will have a dual $\{e^{a}\}$ whose members are such that
\be
e^{a}(e_b) = \delta^a_b.
\label{OrtoLiFra}
\ee
Notice that, on a general manifold, vector fields are (like coordinate systems) only locally defined --- and linear frames, as sets of four such fields, defined only on restricted domains.

These frame fields are the general linear bases on the spacetime differentiable manifold  ${\mathbb R}^{\,3,1}$. The whole set of such bases, under conditions making of it also a differentiable manifold, constitutes the\index{Frame bundle} {\em bundle of linear frames}. A frame field provides, at each point $p \in {\mathbb R}^{\,3,1}$, a basis for the vectors on the tangent space ${T}_p{\mathbb R}^{\,3,1}$. Of course, on the common domains they are defined, each member of a given basis can be written in terms of the members of any other. For example,  \be
e_a = e_a{}^\mu \, \partial_\mu \quad \mbox{and} \quad e^{a} = e^{a}{}_\mu \, dx^\mu,
\ee
and conversely,
\be
\partial_\mu = e^a{}_\mu \, e_a \quad \mbox{and} \quad dx^\mu = e_{a}{}^\mu \, e^{a}.
\label{eq:partialmu}
\ee
On account of the orthogonality conditions (\ref{OrtoLiFra}), the frame components satisfy 
\begin{equation}
e^{a}{}_{\mu} e_{a}{}^{\nu} = \delta_{\mu}^{\nu} \quad \mbox{and} \quad
e^{a}{}_{\mu} e_{b}{}^{\mu} = \delta^{a}_{b}.
\label{eq:frameprops1}
\end{equation}
Notice that these frames, with their bundles, are constitutive parts of spacetime: they are automatically present as soon as spacetime is taken to be a differentiable manifold.

A general linear basis $\{e_{a}\}$ satisfies the commutation relation
\begin{equation}
[e_{a}, e_{b}] = f^{c}{}_{a b} \; e_{c},
\label{eq:comtable0}
\end{equation}
with $f^{c}{}_{a b}$ the so-called structure coefficients, or coefficients of anholonomy, or still the anholonomy of frame $\{e_{a}\}$.\index{Coefficient of anholonomy}
As a simple computation shows, they are defined by
\begin{equation}
f^c{}_{a b} = e_a{}^{\mu} e_b{}^{\nu} (\partial_\nu
e^c{}_{\mu} - \partial_\mu e^c{}_{\nu} ).
\label{fcab0}
\end{equation}
A preferred class is that of inertial frames, denoted $e'_a$, those for which
\be
f'^{a}{}_{cd} = 0.
\label{fcabinertial}
\ee
Such bases $\{e'^{a}\}$ are said to be {\em holonomic}. Of course, all coordinate bases are holonomic. This is not a local property, in the sense that it is valid everywhere for frames belonging to this inertial class.

Consider now the Minkowski spacetime metric, which in cartesian coordinates $\{\bar{x}^\mu \}$ has the form
\be
\bar \eta_{\mu \nu} = \mathrm{diag}(+1,-1,-1,-1).
\label{eq:etaofMinkoST}
\ee
In any other coordinate system, $\eta_{\mu \nu}$ will be a function of the spacetime coordinates. The linear frame
\be
e_{a} = e_{a}{}^{\mu} \, {\partial_{\mu}},
\ee
provides a relation between the tangent-space metric $\eta_{a b}$ and the spacetime metric $\eta_{\mu \nu}$. Such relation is given by
\begin{equation}
\eta_{a b} = {\eta}_{\mu \nu} \, e_{a}{}^{\mu} e_{b}{}^{\nu}.
\label{gtoeta}
\end{equation}
Using the orthogonality conditions (\ref{eq:frameprops1}), the inverse relation is found to be
\begin{equation}
{\eta}_{\mu \nu} = \eta_{a b} \, e^{a}{}_{\mu} e^{b}{}_{\nu}.
\label{eq:tettomet0}
\end{equation}
Independently of whether $e_{a}$ is holonomic or not, or equivalently, whether they are inertial or not, they always relate the tangent Minkowski space to a Minkowski spacetime. These are the frames appearing in special relativity, and are usually called trivial frames --- or trivial tetrads.

\subsubsection{Nontrivial Frames}
\label{sec:tetrads}  \index{Tetrad field!nontrivial}

Nontrivial frames, or nontrivial tetrads, will be denoted by
\be
\{h_{a}\} \quad \mbox{and} \quad \{h^{a}\}.
\ee
They are defined as linear frames whose coefficient of anholonomy is related to both inertial effects {\it and} gravitation. Let us consider a general pseudo-riemannian spacetime with metric components $g_{\mu \nu}$ in some dual holonomic basis $\{d x^{\mu}\}$. The tetrad field
\be
h_{a} = h_{a}{}^{\mu} \, {\partial_{\mu}} \quad \mbox{and} \quad h^a = h^a{}_\mu dx^\mu,
\ee
is a linear basis that relates $g_{\mu \nu}$ to the tangent-space metric $\eta_{a b}$ through the relation
\begin{equation}
\eta_{a b} = g_{\mu \nu} \, h_{a}{}^{\mu} h_{b}{}^{\nu}.
\label{eq:gtoeta}
\end{equation}
The components of the dual basis members
$h^{a} = h^{a}{}_{\nu} dx^{\nu}$ satisfy
\begin{equation}
h^{a}{}_{\mu} \, h_{a}{}^{\nu} = \delta_{\mu}^{\nu} \quad {\rm and} \quad
h^{a}{}_{\mu} \, h_{b}{}^{\mu} = \delta^{a}_{b},
\label{eq:tetradprops1}
\end{equation}
so that Eq.~(\ref{eq:gtoeta}) has the inverse\index{Tetrad field!and metric tensor}
\begin{equation}
g_{\mu \nu} = \eta_{a b} \, h^{a}{}_{\mu} h^{b}{}_{\nu}.
\label{eq:tettomet}
\end{equation}
It follows from these relations that
\be
h \equiv \det (h^a{}_\mu) = \sqrt{-g} \, ,
\ee
with $g = \det(g_{\mu \nu})$. 

A tetrad basis $\{h_{a}\}$ satisfies the commutation relation
\begin{equation}
[h_{a}, h_{b}] = f^{c}{}_{a b}\, h_{c},
\label{eq:comtable}
\end{equation}
with $f^{c}{}_{a b}$ the structure coefficients, or coefficients of anholonomy, of frame $\{h_{a}\}$. The basic difference in relation to the linear bases $\{e_a\}$ is that now the $f^{c}{}_{a b}$ represent both inertial effects and gravitation, and are given by
\index{Coefficient of anholonomy}
\begin{equation}
f^c{}_{a b} = h_a{}^{\mu} h_b{}^{\nu} (\partial_\nu
h^c{}_{\mu} - 
\partial_\mu h^c{}_{\nu} ).
\label{fcab}
\end{equation}
Although nontrivial tetrads are, by definition, anholonomic due to the presence of gravitation, it is still possible that {\em locally}, $f^{c}{}_{a b}$ = $0$. In this case, $d h^{a} = 0$, which means that $h^{a}$ is locally a closed differential form. In fact, if this holds at a point $p$, then there is a neighborhood around $p$ on which functions (coordinates) $x^a$ exist such that
\[
h^{a} = dx^a.
\]
We say that a closed differential form is always locally integrable, or exact. This is the case of locally inertial frames, which are always holonomic. In these frames, inertial effects locally compensate for gravitation.

\subsection{Lorentz Connections}
\label{sec:connections}\index{Lorentz connection}

A {\em Lorentz connection} $A_\mu$, frequently referred to also as {\it spin connection}, is a 1-form assuming values in the Lie algebra of the Lorentz group,
\be
A_\mu = \onehalf \, A^{ab}{}_\mu \, S_{ab},
\ee
with $S_{ab}$ a given representation of the Lorentz generators. As these generators are antisymmetric in the algebraic indices, $A^{ab}{}_\mu$ must be equally antisymmetric in order to be lorentzian. This connection defines the Fock-Ivanenko covariant derivative \cite{fi1,fi2}
\be
\D_\mu = \partial_\mu -  \ihalf \, A^{ab}{}_\mu \, S_{ab},
\label{eq:FockIvanenko} 
\ee
\index{Fock-Ivanenko derivative}whose second part acts only on algebraic, tangent space indices. For a Lorentz vector field $\phi^c$, for example, the representation of the Lorentz generators are matrices $S_{ab}$ with entries \cite{ramond1} 
\be
(S_{ab})^c{}_d = i \left(\eta_{bd} \, \delta_a^c  - \eta_{ad} \, \delta_b^c \right).\label{eq:vecrep}
\ee
The Fock-Ivanenko derivative is, in this case,
\be
\D_\mu \phi^c = \partial_\mu \phi^c + A^{c}{}_{d \mu} \, \phi^d.
\label{VectorFI}
\ee

On account of the soldered character of the tangent bundle, a tetrad field relates tangent space (or internal) tensors with spacetime (or external) tensors. For example, if $\phi^a$ is an internal, or Lorentz vector, then 
\be
\phi^\rho = h_a{}^\rho \, \phi^a
\label{ixe}
\ee
will be a spacetime vector. Conversely, we can write
\be
\phi^a = h^a{}_\rho \, \phi^\rho.
\label{exi}
\ee
On the other hand, due to its non-tensorial character, a connection will acquire a vacuum, non-homogeneous term, under the same operation,
\be
\Gamma^{\rho}{}_{\nu \mu} = h_{a}{}^{\rho} \partial_{\mu} h^{a}{}_{\nu} +
h_{a}{}^{\rho} A^{a}{}_{b \mu} h^{b}{}_{\nu} \equiv
h_{a}{}^{\rho} \, \D_{\mu} h^{a}{}_{\nu},
\label{geco}
\ee
where $\D_{\mu}$ is the covariant derivative (\ref{VectorFI}), in which the generators act on internal (or tangent space) indices only. The inverse relation is, consequently,
\be
A^{a}{}_{b \mu} =
h^{a}{}_{\rho} \partial_{\mu}  h_{b}{}^{\rho} +
h^{a}{}_{\rho} \Gamma^{\rho}{}_{\nu \mu} h_{b}{}^{\nu} \equiv
h^{a}{}_{\rho} \nabla_{\mu}  h_{b}{}^{\rho},
\label{gsc}
\ee
where $\nabla_{\mu}$ is the standard covariant derivative in the connection
$\Gamma^{\nu}{}_{\rho \mu}$, which acts on external indices only. For a spacetime vector $\phi^\nu$, for example, it has the form
\be
\nabla_\mu \phi^\nu = \partial_\mu \phi^\nu + \Gamma^\nu{}_{\rho \mu} \, \phi^\rho.
\ee
Using relations (\ref{ixe}) and (\ref{exi}), it is easy to verify that \cite{kibble}
\be
\D_\mu \phi^d = h^d{}_\rho \, \nabla_\mu \phi^\rho.
\label{iDxeD}
\ee
Equations (\ref{geco}) and (\ref{gsc}) are simply different ways of expressing the property
that the total covariant derivative of the tetrad --- that is, a covariant derivative with connection terms for both internal and external indices --- vanishes identically:
\be
\partial_{\mu} h^{a}{}_{\nu} - \Gamma^{\rho}{}_{\nu \mu} h^{a}{}_{\rho} +
A^{a}{}_{b \mu} h^{b}{}_{\nu} = 0.
\label{todete}
\ee

\subsubsection{Behavior under Lorentz Transformations}
\label{sec:LorentzTransf}
 
A local Lorentz transformation is a transformation of the tangent space coordinates $x^a$:
\be
x'^a = \Lambda^a{}_b(x) \, x^b.
\ee
Under such a transformation, the tetrad transforms according to
\be
h'^{a} = \Lambda^{a}{}_b(x) \, h^b.
\ee
At each point of a riemannian spacetime, Eq.~(\ref{eq:tettomet}) only determines the tetrad up to transformations of the six-parameter Lorentz group in the tangent space indices. This means that there exists actually an infinity of tetrads $h_{a}{}^{\mu}$, each one relating the spacetime metric $g_{\mu \nu}$ to the tangent space metric $\eta_{c d}$. This means that any other Lorentz-rotated tetrad $\{h'_{a}\}$ will also relate the same metrics
\begin{equation}
g_{\mu \nu} = \eta_{c d}\, 
h'^{c}{}_{\mu} h'^{d}{}_{\nu}. 
\label{etatogmunu}
\end{equation} 
Under a local Lorentz transformation $\Lambda^{a}{}_{b}(x)$, the spin connection undergoes the transformation
\be
A'^{a}{}_{b \mu} = \Lambda^{a}{}_{c}(x) \, A^{c}{}_{d \mu} \,
\Lambda_{b}{}^{d}(x) +
\Lambda^{a}{}_{c}(x) \, \partial_{\mu} \Lambda_{b}{}^{c}(x).
\label{ltsc}
\ee
The last, non-homogeneous term appears due to the non-tensorial character of connections.

\subsection{Curvature and Torsion}
\label{sec:CurvTor}

Curvature and torsion require a Lorentz connection to be defined \cite{koba}. Given a Lorentz connection $A^{a}{}_{b \mu}$, the corresponding curvature is a 2-form assuming values in the Lie algebra of the Lorentz group,
\be
R_{\nu \mu} = \onehalf R^{ab}{}_{\nu \mu} \, S_{ab}.
\ee
Torsion is also a 2-form, but assuming values in the Lie algebra of the translation group, 
\be
T_{\nu \mu} = T^{a}{}_{\nu \mu} \, P_a,
\ee
with $P_a = \partial_a$ the translation generators. The curvature and torsion components are given, respectively, by\index{Curvature tensor}
\be
R^{a}{}_{b \nu \mu} = \partial_{\nu} A^{a}{}_{b \mu} -
\partial_{\mu} A^{a}{}_{b \nu} + A^a{}_{e \nu} A^e{}_{b \mu}
- A^a{}_{e \mu} A^e{}_{b \nu}
\label{curvaDef}
\ee
and\index{Torsion tensor}
\be
T^a{}_{\nu \mu} = \partial_{\nu} h^{a}{}_{\mu} -
\partial_{\mu} h^{a}{}_{\nu} + A^a{}_{e \nu} h^e{}_{\mu}
- A^a{}_{e \mu} h^e{}_{\nu}.
\label{tordef}
\ee
Through contraction with tetrads, these tensors can be written in spacetime-indexed forms:
\be
R^\rho{}_{\lambda\nu\mu} = h_a{}^\rho \, h^b{}_\lambda \, R^a{}_{b \nu \mu},
\label{RatoRmi}
\ee
and
\be
T^\rho{}_{\nu \mu} = h_a{}^\rho \, T^a{}_{\nu \mu}.
\label{TatoTmi}
\ee
Using relation (\ref{gsc}), their components are found to be 
\be
\label{sixbm}
R^\rho{}_{\lambda\nu\mu} = \partial_\nu \Gamma^\rho{}_{\lambda \mu} -
\partial_\mu \Gamma^\rho{}_{\lambda \nu} +
\Gamma^\rho{}_{\eta \nu} \Gamma^\eta{}_{\lambda \mu} -
\Gamma^\rho{}_{\eta \mu} \Gamma^\eta{}_{\lambda \nu}
\ee
and
\be
T^\rho{}_{\nu \mu} =
\Gamma^\rho{}_{\mu\nu} - \Gamma^\rho{}_{\nu\mu}.
\label{sixam}
\ee

Since the spin connection $A^a{}_{b \nu}$ is a four-vector in the last index, it satisfies
\be
A^a{}_{bc} = A^a{}_{b \nu} \, h_c{}^\nu.
\ee
It can thus be verified that, in the anholonomic basis $\{h_a\}$, the curvature and
torsion components are given respectively by 
\be
R{}^{a}{}_{b cd } =    
h_c \left(A^a{}_{b d} \right) - h_d \left(A^a{}_{b c} \right) + A^a{}_{e c}A^e{}_{b d} - A^a{}_{e d} A^e{}_{b c} - f^{e}{}_{c d} A^a{}_{b e}  
\label{13bm}
\ee
and 
\be\
T^a{}_{bc} = A^a{}_{cb} - A^a{}_{bc} - f^a{}_{bc}, 
\label{13am}
\ee
where, we recall, $h_c = h_c{}^\mu \partial_\mu$. Use of~(\ref{13am}) for three different combinations of indices gives
\begin{equation}%
A^{a}{}_{b c} = \onehalf (f_{b}{}^{a}{}_{c} + T_{b}{}^{a}{}_{c} 
+ f_{c}{}^{a}{}_{b} + T_{c}{}^{a}{}_{b}
- f^{a}{}_{b c} - T^{a}{}_{b c}).
\label{tobetaken2}
\end{equation}%
This expression can be rewritten in the form
\be
A^a{}_{bc} = \Abol^a{}_{bc} + K^a{}_{bc},
\label{rela0alge}
\ee
where
\begin{equation}%
\Abol^{a}{}_{b c} = \onehalf \left(f_{b}{}^{a}{}_{c} 
+ f_{c}{}^{a}{}_{b} - f^{a}{}_{b c} \right) 
\label{tobetaken30}
\end{equation}%
is the usual expression of the general relativity spin connection in terms of the coefficients of anholonomy, and\index{Lorentz connection!of general relativity}
\begin{equation}%
K^{a}{}_{b c} = \onehalf \left(T_{b}{}^{a}{}_{c} 
+ T_{c}{}^{a}{}_{b} - T^{a}{}_{b c} \right) 
\label{contorDef}
\end{equation}%
is the {contortion tensor}.\index{Contortion tensor}
The corresponding expression in terms of the spacetime-indexed linear connection reads
\be
\Gamma^\rho{}_{\mu\nu} = {\stackrel{\circ}{\Gamma}}{}^{\rho}{}_{\mu \nu} +
K^\rho{}_{\mu\nu},
\label{prela0}
\ee
where
\be
{\stackrel{\circ}{\Gamma}}{}^{\sigma}{}_{\mu \nu} = \onehalf \,
g^{\sigma \rho} \left( \partial_{\mu} g_{\rho \nu} +
\partial_{\nu} g_{\rho \mu} - \partial_{\rho} g_{\mu \nu} \right)
\label{lci}
\ee
is the zero-torsion Christoffel, or Levi-Civita connection, and\index{Christoffel connection} \index{Levi-Civita connection} 
\be
K^\rho{}_{\mu\nu} = {\textstyle
\frac{1}{2}} \left(T_\nu{}^\rho{}_\mu+T_\mu{}^\rho{}_\nu-
T^\rho{}_{\mu\nu}\right)
\label{contor}
\ee
is the spacetime-indexed contortion tensor. Equations~(\ref{rela0alge}) and (\ref{prela0}) are actually the content of a theorem, which states that any Lorentz connection can be decomposed into the spin connection of general relativity plus the contortion tensor \cite{koba}. As is well-known, the Levi-Civita connection of a general spacetime metric has vanishing torsion, but non-vanishing curvature:
\be
\Tbol^\rho{}_{\nu \mu} = 0 \quad \mbox{and} \quad \Rbol^{\rho}{}_{\lambda \nu \mu} \neq 0.
\ee

\subsection{Purely Inertial Lorentz Connection}
\label{InerGra}

In special relativity, Lorentz connections represent inertial effects present in a given frame. In order to obtain the explicit form of such connections, let us recall that the class of inertial (or holonomic) frames, denoted by $e'^a{}_\mu$, is defined by all frames for which $f'^{c}{}_{a b} = 0$. In a general coordinate system, the frames belonging to this class have the holonomic form
\be
e'^a{}_\mu = \partial_\mu x'^a,
\label{frame0}
\ee
with $x'^a$ a spacetime-dependent Lorentz vector: $x'^a = x'^a(x^\mu)$. Under a local Lorentz transformation, the holonomic frame (\ref{frame0}) transforms according to
\be
e^a{}_\mu = \Lambda^a{}_b(x) \, e'^b{}_\mu.
\label{LoreTrans-e}
\ee
As a simple computation shows, it has the explicit form
\be
e^a{}_\mu = \partial_\mu x^a + \Aw^a{}_{b \mu} \, x^b \equiv \Dw_\mu x^a,
\label{InertiaTetrad}
\ee
where
\be
\Aw^a{}_{b \mu} = \Lambda^a{}_e(x) \, \partial_\mu \Lambda_b{}^e(x)
\label{InerConn}
\ee
is a Lorentz connection that represents the inertial effects present in the new frame $e^a{}_\mu$.\index{Lorentz connection!purely inertial} As can be seen from Eq.~(\ref{ltsc}), it is just the connection obtained from a Lorentz transformation of the vanishing spin connection $\Aw'^e{}_{d \mu} = 0$:
\be
\Aw^a{}_{b \mu} = \Lambda^a{}_e(x) \, \Aw'^e{}_{d \mu} \, \Lambda_b{}^d(x) +
\Lambda^a{}_e(x) \, \partial_\mu \Lambda_b{}^e(x).
\ee
Starting from an inertial frame, different classes of frames are obtained by performing {\em local} (point-dependent) Lorentz transformations $\Lambda^{a}{}_b(x^\mu)$. Within each class, the infinitely many frames are related through {\em global} (point-independent) Lorentz transformations, $\Lambda^{a}{}_b =$~constant.

Each component of the inertial connection~(\ref{InerConn}), which is sometimes referred to as the Ricci coefficient of rotation \cite{MTW732}, represents a different inertial effect \cite{papini}. Owing to its presence, the transformed frame $e^a{}_\mu$ is no longer holonomic. In fact, its coefficient of anholonomy is given by
\be
f^c{}_{a b} = - \left(\Aw^c{}_{a b} - \Aw^c{}_{b a} \right),
\ee
where we have used the identity $\Aw^a{}_{b c} = \Aw^a{}_{b \mu} \, e_c{}^\mu$. Of course, as a purely inertial connection, $\Aw^a{}_{b \mu}$ has vanishing curvature and torsion:
\be
\Rw^{a}{}_{b \nu \mu} \equiv \partial_{\nu} \Aw^{a}{}_{b \mu} -
\partial_{\mu} \Aw^{a}{}_{b \nu} + \Aw^a{}_{e \nu} \, \Aw^e{}_{b \mu}
- \Aw^a{}_{e \mu} \, \Aw^e{}_{b \nu} = 0
\label{curvaDefW}
\ee
and\index{Torsion tensor}
\be
\Tw^a{}_{\nu \mu} \equiv \partial_{\nu} e^{a}{}_{\mu} -
\partial_{\mu} e^{a}{}_{\nu} + \Aw^a{}_{e \nu} \, e^e{}_{\mu}
- \Aw^a{}_{e \mu} \, e^e{}_{\nu} = 0.
\label{tordefW}
\ee

\subsection{Equation of Motion of Free Particles}

To see how a purely inertial connection shows up in a concrete example, let us consider the equation of motion of a free particle. In the class of inertial frames $e'^a{}_\mu$, such particle is described by the equation of motion
\be
\frac{d u'^a}{d\sigma} = 0,
\ee
with $u'^a$ the anholonomic four-velocity, and
\be
d \sigma^2 = \eta_{\mu \nu} \, dx^\mu dx^\nu
\label{MinkoInter}
\ee
the quadratic Minkowski interval. In an anholonomic frame $e^a{}_\mu$, related to $e'^a{}_\mu$ by the local Lorentz transformation (\ref{LoreTrans-e}), the equation of motion assumes the manifestly covariant form under local Lorentz transformations
\be
\frac{d u^a}{d\sigma} + \Aw^a{}_{b \mu} \, u^b \, u^\mu = 0,
\label{anholoEM}
\ee
where
\be
u^a = \Lambda^a{}_b(x) \, u'^b
\ee
is the Lorentz transformed four-velocity, and
\be
u^\mu = u^a \, e_a{}^\mu
\label{uea}
\ee
is the usual, holonomic four-velocity
\be
u^\mu = \frac{d x^\mu}{d\sigma}.
\ee
Observe that the inertial forces coming from the frame non-inertiality are represented by the inertial connection on the left-hand side of the equation~(\ref{anholoEM}), which is non-covariant by its very nature. Observe also that it is invariant under general coordinate transformations.

In terms of the holonomic four-velocity written in cartesian coordinates $\{\bar{x}^\mu\}$, the particle equation of motion has the form
\be
\frac{d \bar{u}^\rho}{d\sigma} = 0.
\ee
Under a general coordinate transformation $\bar{x}^\mu \to x^\mu$, it assumes the manifestly covariant form under general coordinate transformations
\be
\frac{d u^\rho}{d\sigma} + \gammaw^\rho{}_{\nu \mu} \, u^\nu u^\mu = 0,
\label{holoEM}
\ee
where \cite{mosna}
\be
\gammaw^\rho{}_{\nu \mu} = \onehalf \eta^{\rho \lambda} \left(\partial_\nu \eta_{\lambda \mu}
+ \partial_\mu \eta_{\lambda \nu} - \partial_\lambda \eta_{\nu \mu} \right)
\label{InerConn2}
\ee
is a flat, coordinate-related connection, with $\eta_{\nu \mu}$ the Minkowski metric written in the general coordinate system $\{x^\mu\}$. Of course, since the equations of motion (\ref{anholoEM}) and (\ref{holoEM}) describe the same free particle, they are equivalent ways of writing the same equation of motion. This means that connections (\ref{InerConn}) and (\ref{InerConn2}) are different ways of writing the very same inertial connection. In fact, using relation (\ref{uea}), it is an easy task to verify that they are related by
\be
\Aw^a{}_{b \mu} =
e^a{}_\rho \partial_\mu e_b{}^\rho + e^a{}_\rho \gammaw^\rho{}_{\nu \mu} \, e_b{}^\nu \equiv
e^a{}_\rho \, \nablaw_\mu e_b{}^\rho,
\label{STindIneCon2}
\ee
which is a relation of the form (\ref{gsc}) between equivalent connections. We can then conclude that local Lorentz transformations are equivalent to general coordinate transformations in the sense that they give rise to the very same inertial connection. In Section~\ref{GaugeGrav} we will discuss further the implications of this equivalence for gravitation.

\section{Teleparallel Gravity: A Brief Review}
\label{sec:TeleGrav}\index{Teleparallel gravity!a r\'esum\'e of}

For the sake of completeness we present in this section a short review of teleparallel gravity, as well as discuss its equivalence to general relativity.

\subsection{Translational Gauge Potential}
\label{sec:BasicFields}
\index{Teleparallel gravity!as a gauge theory}

Teleparallel gravity corresponds to a gauge theory for the translation group \cite{livro2}. As such, the gravitational field is represented by a translational gauge potential ${B^a{}_\mu}$, a 1-form assuming values in the Lie algebra of the translation group,\index{Teleparallel gravity!fundamental field}
\be
B_\mu = B^a{}_\mu \, P_a,
\ee
with $P_a = \partial_a$ the translation generators. On account of the translational coupling prescription, it appears as the non-trivial part of the tetrad,
\be
h^a{}_\mu = e^a{}_\mu + B^a{}_\mu,
\label{TeleTetrada2}
\ee
where
\be
e^a{}_\mu \equiv \Dw_\mu x^a = \partial_\mu x^a + \Aw^a{}_{b \mu} \, x^b
\label{NonGraTetra}
\ee
is the trivial (non-gravitational) tetrad (\ref{InertiaTetrad}). Under an infinitesimal gauge translation
\be
\delta x^a = \varepsilon^b P_b \, x^a \equiv \varepsilon^a,
\ee
with $\varepsilon^a \equiv \varepsilon^a(x^\mu)$ the transformation parameters, the gravitational potential $B^a{}_\mu$ transforms according to
\be
\delta B^a{}_\mu = - \, \Dw_\mu \varepsilon^a.
\label{BamGauTrans}
\ee
The tetrad (\ref{TeleTetrada2}) is consequently gauge invariant:
\be
\delta h^a{}_\mu = 0.
\ee
This is a matter of consistency as a gauge transformation cannot change the spacetime metric.

\subsection{Teleparallel Spin Connection}
\label{SpinConne}
\index{Lorentz connection!of teleparallel gravity}

The gravitational field in teleparallel gravity is fully represented by the translational gauge potential ${B^a{}_\mu}$. This means that in this theory Lorentz connections keep their special-relativistic role of representing inertial effects only. The fundamental Lorentz connection of teleparallel gravity is consequently the purely inertial connection (\ref{InerConn}), which has of course vanishing curvature:
\be
\Rw^{a}{}_{b \mu \nu} = \partial_{\mu} \Aw^{a}{}_{b \nu} -
\partial_{\nu} \Aw^{a}{}_{b \mu} + \Aw^a{}_{e \mu} \Aw^e{}_{b \nu}
- \Aw^a{}_{e \nu} \Aw^e{}_{b \mu} = 0.
\ee
For a tetrad involving a non-trivial translational gauge potential, that is, for
\be
B^a{}_\mu \neq \Dw_\mu \varepsilon^a,
\label{NonTriB}
\ee
its torsion will be non-vanishing:
\be
\Tw^a{}_{\mu \nu} = \partial_{\mu} h^{a}{}_{\nu} -
\partial_{\nu} h^{a}{}_{\mu} + \Aw^a{}_{e \mu} h^e{}_{\nu}
- \Aw^a{}_{e \nu} h^e{}_{\mu} \neq 0.
\ee
Using the trivial identity
\be
\Dw_\mu \Dw_\nu x^a - \Dw_\nu \Dw_\mu x^a = 0,
\ee
it can be rewritten in the form
\be
\Tw^a{}_{\mu \nu} = \partial_\mu B^a{}_\nu - \partial_\nu B^a{}_\mu +
\Aw^a{}_{b \mu} B^b{}_{\nu} - \Aw^a{}_{b \nu} B^b{}_{\mu},
\label{tfs}
\ee
which is the field strength of teleparallel gravity. In this theory, therefore, gravitation is represented by torsion, not by curvature. On account of the gauge invariance of the tetrad, the field strength is also invariant under gauge transformations:
\be
\Tw'^a{}_{\mu \nu} = \Tw^a{}_{\mu \nu}.
\ee
This is actually an expected result. In fact, considering that the generators of the adjoint representation are the coefficients of structure of the group taken as matrices, and considering that these coefficients vanish for abelian groups, fields belonging to the adjoint representation of abelian gauge theories will always be gauge invariant --- a well-known property of electromagnetism.

The spacetime linear connection corresponding to the inertial spin connection (\ref{InerConn}) is
\be
\Gammaw^{\rho}{}_{\nu \mu} = h_{a}{}^{\rho} \partial_{\mu} h^{a}{}_{\nu} +
h_{a}{}^{\rho}\Aw^a{}_{b \mu} \, h^b{}_\nu \equiv
h_{a}{}^{\rho} \, \Dw_{\mu} h^{a}{}_{\nu}.
\label{gecow}
\ee
This is the so-called Weitzenb\"ock connection.\index{Weitzenb\"ock connection} Its definition is equivalent to the identity
\be
\partial_{\mu}h^a{}_{\nu} +
\Aw^a{}_{b \mu} \, h^b{}_\nu -
\Gammaw^{\rho}{}_{\nu \mu} \, h^a{}_{\rho} = 0.
\label{cacd1}
\ee
In the class of frames in which the spin connection $\Aw^a{}_{b \mu}$ vanishes, it reduces to
\be
\partial_{\mu}h^a{}_{\nu} -
\Gammaw^{\rho}{}_{\nu \mu} \, h^a{}_{\rho} = 0,
\label{cacd0}
\ee
which is the so-called absolute, or distant parallelism condition, from where teleparallel gravity got its name. It is important to remark that, at the time the term absolute, or distant parallelism condition was coined, no one was aware that this condition holds only on a very specific class of frames. The general expression valid in any frame is that given by Eq.~(\ref{cacd1}). This means essentially the the tetrad is not actually parallel-transported everywhere by the Weitzenb\"ock connection. The name ``teleparallel gravity'' is consequently not appropriate. Of course, for historical reasons we shall keep it.
\index{Absolute parallelism condition}\index{Distant parallelism condition}
\index{Teleparallel gravity!origin of name}

\subsection{Teleparallel Lagrangian}
\index{Teleparallel gravity!lagrangian}

As a gauge theory for the translation group, the action functional of teleparallel gravity can be written in the form \cite{fadnov}
\begin{equation}
\Sw = \frac{1}{2ck} \int \, \eta_{ab} \, \Tw^a \wedge {\star}\Tw^b,
\label{action1}
\end{equation}
where
\begin{equation}
\Tw^a = \textstyle{\frac{1}{2}} \, \Tw^a{}_{\mu\nu} \, d x^\mu \wedge d x^\nu 
\label{Tform}
\end{equation}
is the torsion 2-form, ${\star}\Tw^a$ is the corresponding dual form, and $k = 8 \pi G/c^4$. More explicitly,
\begin{equation}
\Sw =
\frac{1}{8 c k} \int \, \eta_{ab} \, \Tw^a{}_{\mu\nu} \;
\star\Tw^b{}_{\rho \sigma} \, d x^\mu \wedge d x^\nu \wedge d x^\rho \wedge d x^\sigma.
\label{action3}
\end{equation}
Taking into account the identity
\begin{equation}
d x^\mu \wedge d x^\nu \wedge d x^\rho \wedge d x^\sigma = - \,
\epsilon^{\mu \nu \rho \sigma} \, h \, d^4x,
\end{equation}
with $h = \det (h^a{}_\mu)$, the action functional assumes the form
\begin{equation}
\Sw = -
\frac{1}{8 c k} \int \, \Tw_{a\mu\nu}
\star\Tw^{a}{}_{\rho \sigma} \, \epsilon^{\mu \nu \rho \sigma} \; h \, d^4x.
\label{action4}
\end{equation}
Using then the generalized dual definition for soldered bundles \cite{SolDual}
\be
\star T^a{}_{\mu \nu} = \frac{h}{2} \, \epsilon_{\mu \nu \alpha \beta} \, S^{a \alpha \beta},
\label{galstar4}
\ee
it reduces to
\begin{equation}
\Sw =
\frac{1}{4ck} \int \Tw^a{}_{\rho \sigma} \,
\sw_a{}^{\rho \sigma} \, h \, d^4x,
\label{TeleAction}
\end{equation}
where
\be
\sw_a{}^{\rho \sigma} \equiv - \,
\sw_a{}^{\sigma \rho} = h_a{}^\nu \left( \Kw^{\rho \sigma}{}_{\nu} - \delta_\nu{}^{\sigma} \;  \Tw^{\theta \rho}{}_{\theta} + \delta_\nu{}^{\rho} \; \Tw^{\theta \sigma}{}_{\theta} \right)
\label{supote}
\ee
is the superpotential, with
\be
\Kw^{\rho \sigma}{}_{\nu} = {\textstyle \frac{1}{2}} \left( \Tw^{\sigma\rho}{}_{\nu}
+ \Tw_{\nu}{}^{\rho \sigma} - \Tw^{\rho \sigma}{}_{\nu} \right)
\ee
the contortion of the teleparallel torsion.\index{Contortion tensor}
The lagrangian corresponding to the above action is \cite{maluf94}
\be
\Lw =
\frac{h}{4k} \, \Tw_{a \mu\nu} \,
\sw^{a \mu\nu}.
\label{TeleLa}
\ee
Using relation (\ref{prela0}) for the specific case of teleparallel torsion, it is possible to show that
\begin{equation}
\Lw = \Lbol - \partial_\mu \left(2 \, h \, k^{-1} \,
\Tw^{\nu \mu}{}_\nu \right),
\label{LagraEquiva}
\end{equation}
where
\begin{equation}
\Lbol = - \, \frac{\sqrt{-g}}{2 k} \; \Rbol
\label{e-hl}
\end{equation}
is the Einstein-Hilbert lagrangian of general relativity. Up to a divergence, therefore, the teleparallel lagrangian is equivalent to the lagrangian of general relativity.\index{Teleparallel gravity!equivalence with general relativity}

One may wonder why the lagrangians are equivalent up to a divergence term. To understand that, let us recall that the Einstein-Hilbert lagrangian (\ref{e-hl}) depends on the tetrad, as well as on its first and second derivatives. The terms containing second derivatives, however, reduce to a divergence term \cite{landau}. In consequence, it is possible to rewrite the Einstein-Hilbert lagrangian in a form stating this aspect explicitly, 
\be
\Lbol = \Lbol_1 + \partial_\mu (\sqrt{-g} \, w^\mu),
\ee 
where $\Lbol_1$ is a lagrangian that depends solely on the tetrad and its first derivatives, and $w^\mu$ is a four-vector. On the other hand, the teleparallel lagrangian (\ref{TeleLa}) depends only on the tetrad and its first derivative. The divergence term in the equivalence relation (\ref{LagraEquiva}) is then necessary to account for the different orders of the teleparallel and the Einstein-Hilbert lagrangians. We mention in passing that in classical field theory the lagrangians involve only the field and its first derivative. We can then say that teleparallel gravity is more akin to a field theory than general relativity. In Section~\ref{GenuConne} this point will be discussed in further details.

\subsection{Field Equations}
\label{sec:FieldEquations}
 \index{Teleparallel gravity!field equation}

Consider the lagrangian
\begin{equation}
{\mathcal L} = \Lw + {\mathcal L}_s,
\end{equation}
with ${\mathcal L}_s$ the lagrangian of a general source field. Variation with respect to the gauge potential $B^a{}_\rho$ (or equivalently, in terms of the tetrad $h^a{}_\rho$) yields the teleparallel version of the gravitational field equation
\be
\partial_\sigma(h \sw_a{}^{\rho \sigma}) -
k \, h \jw_{a}{}^{\rho} = k \, h {\Theta}_{a}{}^{\rho}.
\label{tfe10}
\ee
In this equation,  
\be
h \jw_{a}{}^{\rho} \equiv -\,
\frac{\partial \Lw}{\partial h^a{}_{\rho}} =
\frac{1}{k} \, h_a{}^{\mu} \, \sw_c{}^{\nu \rho} \,
\Tw^c{}_{\nu \mu} - \frac{h_a{}^{\rho}}{h} \, \Lw +
\frac{1}{k} \, \Aw^c{}_{a \sigma} \sw_c{}^{\rho \sigma}
\label{ptem10}
\ee
stands for the gauge current, which in this case represents the Noether energy-momentum pseudo-current of gravitation plus inertial effects \cite{gemt}, and
\begin{equation}
h {\Theta}_{a}{}^{\rho} = -\, \frac{\delta {\mathcal L}_s}{\delta h^a{}_{\rho}} \equiv -
\left( \frac{\partial {\mathcal L}_s}{\partial h^a{}_{\rho}} -
\partial_\mu \frac{\partial {\mathcal L}_s}{\partial_\mu \partial h^a{}_{\rho}} \right)
\label{memt1}
\end{equation}
is the source energy-momentum tensor.\index{Energy-momentum tensor!of source fields}
Due to the anti-symmetry of the superpotential in the last two indices, the {total} (gravitational plus inertial plus source) energy-momentum density is conserved in the ordinary sense:
\be
\partial_\rho \big(h \jw_{a}{}^{\rho} + h\, {\Theta}_{a}{}^{\rho} \big) = 0.
\ee

The left-hand side of the gravitational field equation~(\ref{tfe10}) depends on $\Aw^a{}_{b \mu}$ only. Using identity (\ref{rela0alge}) for the specific case of the inertial connection $\Aw^a{}_{b \mu}$,
\be
\Aw^a{}_{b\mu} = \Abol^a{}_{b\mu} + \Kw^a{}_{b\mu},
\label{rela0Tele}
\ee
through a lengthy but straightforward calculation, it can be rewritten in terms of $\Abol^a{}_{b \mu}$ only:
\begin{equation}
\partial_\sigma \big(h \sw_a{}^{\rho \sigma}\big) -
k \, h \jw_{a}{}^{\rho} =
h \big({\stackrel{\circ}{R}}_a{}^{\rho} -
\onehalf \, h_a{}^{\rho} \,
{\stackrel{\circ}{R}} \big).
\label{ident}
\end{equation}
As expected due to the equivalence between the corresponding lagrangians, the teleparallel field equation (\ref{tfe10}) is equivalent to Einstein's field equation\index{Teleparallel gravity!equivalence with general relativity}
\be
\Rbol_a{}^{\rho} -
\onehalf \, h_a{}^{\rho}
\Rbol = k {\Theta}_{a}{}^{\rho}.
\ee
Observe that the energy-momentum tensor appears as the source in both theories: as the source of curvature in general relativity, and as the source of torsion in teleparallel gravity. This is in agreement with the idea that curvature and torsion are related to the same degrees of freedom of the gravitational field.

\section{Achievements of Teleparallel Gravity}

Despite being equivalent to general relativity, teleparallel gravity shows many conceptual distinctive features. In this section we discuss some of these features, as well as explore their possible consequences for the study of both classical and quantum gravity.

\subsection{Separating Inertial Effects from Gravitation}
\label{SepaInerGrav}\index{Separating gravitation and inertia}

In teleparallel gravity, the tetrad field has the form
\be
h^a{}_\mu = \Dw_\mu x^a + B^a{}_\mu.
\label{Tetra6}
\ee
The first term on the right-hand side represents the frame and the inertial effects present on it. The second term, given by the translational gauge potential, represents gravitation only. This means that both inertia and gravitation are included in the tetrad $h^a{}_\mu$. As a consequence, its coefficient of anholonomy,
\begin{equation}
f^c{}_{a b} = h_a{}^{\mu} h_b{}^{\nu} (\partial_\nu
h^c{}_{\mu} - \partial_\mu h^c{}_{\nu}),
\label{fcabBIS}
\end{equation}
will also represent both inertia and gravitation. Of course, the same is true of the spin connection of general relativity,
\begin{equation}%
\Abol^{a}{}_{b \mu} = \onehalf \, h^c{}_\mu \, (f_b{}^a{}_c + f_c{}^a{}_b - f^{a}{}_{b c}). 
\label{tobetaken3}
\end{equation}%
Now, according to the identity (\ref{rela0Tele}), such spin connection can be decomposed in the form
\be
\Abol^{a}{}_{b \mu} = \Aw^{a}{}_{b \mu} - \Kw^{a}{}_{b \mu}.
\label{splittingPartimec}
\ee
Since $\Aw^{a}{}_{b \mu}$ represents inertial effects only, whereas $\Kw^{a}{}_{b \mu}$ represents the gravitational field, the above identity amounts actually to a decomposition of the general relativity spin connection~(\ref{tobetaken3}) into inertial and gravitational parts.

To see that this is in fact the case, let us consider a locally inertial frame in which the spin connection of general relativity vanishes:
\be
\Abol^{a}{}_{b \mu} \doteq 0.
\label{146}
\ee
In such local frame, although present, gravitation becomes locally undetectable. Making use of identity (\ref{splittingPartimec}), the local vanishing of $\Abol^{a}{}_{b \mu}$ can be rewritten in the form
\be
\Aw^{a}{}_{b \mu} \doteq \Kw^{a}{}_{b \mu}.
\ee
This expression shows explicitly that, in such a local frame inertial effects (left-hand side) exactly compensate for gravitation (right-hand side) \cite{Einstein056}. The possibility of separating inertial effects from gravitation is an outstanding property of teleparallel gravity. It opens up many interesting new roads for the study of gravitation, which are not possible in the context of general relativity.

\subsection{Geometry Versus Force}

In general relativity, the trajectories of spinless particles are described by the geodesic equation
\be
\frac{du^a}{ds} + 
\Abol^{a}{}_{b \mu} \, u^b u^{\mu} = 0,
\label{eq:geodesic}
\ee
where $ds^2 = g_{\mu \nu} \, dx^\mu dx^\nu$ is the riemannian spacetime quadratic interval. It says essentially that the four-acceleration of the particle vanishes:
\be
{\stackrel{\circ}{a}}{}^{a} = 0.
\ee
This means that in general relativity {\em there is no the concept of gravitational force}. Using identity (\ref{splittingPartimec}), the geodesic equation can be rewritten in terms of a purely inertial connection and its torsion. The result is\index{Teleparallel gravity!force equation}

\be
\frac{du^a}{ds} +
\Aw^a{}_{b \mu} \, u^b \, u^\mu = \Kw^a{}_{b \mu} \, u^b \, u^\mu.
\label{TeleForceEqu}
\ee
This is the teleparallel equation of motion of a spinless particle as seen from a general Lorentz frame. Of course, it is equivalent to the geodesic equation~(\ref{eq:geodesic}). There are conceptual differences, though. In general relativity, a theory fundamentally based on the equivalence principle, curvature is used to {\it geometrize} the gravitational interaction. The gravitational interaction in this case is described by letting (spinless) particles to follow the curvature of spacetime. Geometry replaces the concept of force, and the trajectories are determined, not by force equations, but by geodesics. Teleparallel gravity, on the other hand, attributes gravitation to torsion, which accounts for gravitation not by geometrizing the interaction, but by acting as a force \cite{AndPer97}. In consequence, there are no geodesics in teleparallel gravity, only force equations similar to the Lorentz force equation of electrodynamics (We remark in passing that this is in agreement with the gauge structure of teleparallel gravity in the sense that gauge theories always describe the classical interaction through a force). Notice that the inertial forces coming from the frame non-inertiality are represented by the connection on the left-hand side, which is non-covariant by its very nature. In teleparallel gravity, therefore, whereas the gravitational effects are described by a covariant force, the inertial effects of the frame remain {\it geometrized} in the sense of general relativity. In the geodesic equation~(\ref{eq:geodesic}), both inertial and gravitational effects are described by the connection term on the left-hand side.

\subsection{Gravitational Energy-Momentum Density}
\label{epi23} \index{Teleparallel gravity!energy localization in}

All fundamental fields have a well-defined local energy-momentum density. It is then expected that the same should happen to the gravitational field. However, no tensorial expression for the gravitational energy-momentum density can be defined in the context of general relativity. The basic reason for this impossibility is that both gravitational and inertial effects are mixed in the spin connection of the theory, and cannot be separated. Even though some quantities, like curvature, are not affected by inertial effects, some others turn out to depend on it. For example, the energy-momentum density of gravitation will necessarily include both the energy-momentum density of gravity and the energy-momentum density of the inertial effects present in the frame. Since the inertial effects are essentially non-tensorial --- they depend on the frame --- the quantity defining the energy-momentum density of the gravitational field in this theory always shows up as a non-tensorial object. Some examples of different pseudotensors can be found in Refs.~\cite{pseudo1, pseudo2, pseudo3, pseudo4, pseudo5, pseudo6, pseudo7, pseudo8, pseudo9}.

On the other hand, owing to the possibility of separating gravitation from inertial effects in teleparallel gravity, it turns out possible to write down an energy-momentum density for gravitation only, excluding the contribution from inertia. Such quantity is a tensorial object. To see how this is possible, let us consider the sourceless version of the teleparallel field equation (\ref{tfe10}),
\be
\partial_\sigma (h \sw_a{}^{\rho \sigma}) -
k \, h \jw_{a}{}^{\rho} = 0,
\label{tfe0}
\ee
where
\be
h \jw_{a}{}^{\rho} =
\frac{1}{k} \, h_a{}^{\mu} \, \sw_c{}^{\nu \rho} \,
\Tw^c{}_{\nu \mu} - \frac{h_a{}^{\rho}}{h} \, \Lw +
\frac{1}{k} \, \Aw^c{}_{a \sigma} \sw_c{}^{\rho \sigma}
\label{ptem10bis}
\ee
is the usual gravitational energy-momentum pseudo-current, which is conserved in the ordinary sense:
\be
\partial_\rho (h \jw_{a}{}^{\rho}) = 0.
\label{iplustCon}
\ee
This is actually a matter of necessity: since the derivative is not covariant, the conserved current cannot be covariant either so that the conservation law itself is covariant --- and consequently physically meaningful.

Using now the fact that the last, non-tensorial term of the pseudo-current (\ref{ptem10bis}) together with the potential term make up a Fock-Ivanenko covariant derivative,
\be
\partial_\sigma (h \sw_a{}^{\rho \sigma}) -
\Aw ^c{}_{a \sigma} (h \, \sw_{c}{}^{\rho \sigma}) \equiv \Dw_\sigma (h \sw_a{}^{\rho \sigma}),
\label{DotFI}
\ee
the field equation (\ref{tfe0}) can be rewritten in the form
\be
\Dw_\sigma (h \sw_a{}^{\rho \sigma}) -
k \, h \, \tw_{a}{}^{\rho} = 0,
\label{fe11}
\ee
where
\be
\tw_{a}{}^{\rho} =
\frac{1}{k} \, h_a{}^{\lambda} \, \sw_c{}^{\nu \rho} \,
\Tw^c{}_{\nu \lambda} - \frac{h_a{}^{\rho}}{h} \, \Lw
\label{graem}
\ee
is a {tensorial current that represents the energy-momentum of gravity alone} \cite{gemt}.\index{Energy-momentum tensor!of gravitation}
Considering that the teleparallel spin connection (\ref{InerConn}) has vanishing curvature, the corresponding Fock-Ivanenko derivative is commutative:
\be
[\Dw_\rho , \Dw_\sigma] = 0.
\ee
Taking into account the anti-symmetry of the su\-per\-potential in the last two indices, it follows from the field equation (\ref{fe11}) that the tensorial current (\ref{graem}) is conserved in the  covariant sense:
\be
\Dw_\rho (h \tw_{a}{}^{\rho}) = 0.
\label{GravEMcon}
\ee
This is again a matter of necessity: a covariant current can only be conserved in the covariant sense. Of course, since it does not represent the total energy-momentum density --- in the sense that the inertial energy-momentum density is not included --- it does not need to be truly conserved. Only the total energy-momentum density $\jw_{a}{}^{\rho}$ must be truly conserved. 

It should be remarked that the use of pseudotensors to compute the energy of a gravitational system requires some amount of handwork to get the physically relevant result. The reason is that, since the pseudotensor includes the contribution from the inertial effects, which is in general divergent for large distances (recall the centrifugal force, for example), the space integration of the energy density usually yields divergent results. It is then necessary to use appropriate coordinates --- like for example cartesian coordinates \cite{CarteCoo} --- or to make use of a regularization process to eliminate the spurious contributions coming from the inertial effects \cite{MalufRegu}. On the other hand, on account of the tensorial character of the teleparallel energy-momentum density of gravity, its use to compute the energy of any gravitational system always gives the physical result, no matter the coordinates or frames used to make the computation, eliminating in this way the necessity of using appropriate coordinates or a regularizing process \cite{ReguRole}.

\subsection{A Genuine Gravitational Variable}
\label{GenuConne}

Due to the fact that the spin connection of general relativity involves both gravitation and inertial effects, it is always possible to find a local frame --- called {\em locally inertial frame} --- in which inertial effects exactly compensate for gravitation, and the connection vanishes at that point:
\be
\Abol^{a}{}_{b \mu} \doteq 0.
\ee
Since there is gravitational field at that point, such connection cannot be considered a genuine gravitational variable in the usual sense of classical field theory. Strictly speaking, therefore, general relativity is not a true field theory. There is an additional problem: the non-covariant behavior of $\Abol^{a}{}_{b \mu}$ under local Lorentz transformations is due uniquely to its inertial content, not to gravitation itself. To see it, consider the decomposition (\ref{splittingPartimec}): whereas the first term on the right-hand side represents its inertial, non-covariant part, the second term represents its gravitational part, which is a tensor. This means that it is not a genuine gravitational connection either, but an inertial connection. 

In teleparallel gravity, on the other hand, the gravitational field is represented by a trans\-lational-valued gauge potential
\be
B_\mu = B^a{}_\mu P_a,
\ee
which shows up as the non-trivial part of the tetrad. Considering that the translational gauge potential represents gravitation only, not inertial effects, it can be considered a true gravitational variable in the sense of classical field theory. Notice, for example, that it is not possible to find a local frame in which it vanishes at a point. Furthermore, it is also a genuine gravitational connection: its connection behavior under gauge translations is related uniquely to its gravitational content. Put together, these properties show that, in contrast to general relativity, teleparallel gravity is a (background-dependent) true field theory.

\subsection{Gravitation and Gauge Theories}
\label{GaugeGrav}
\index{Teleparallel gravity!as a field theory}

If general relativity is not a true field theory, it cannot be a gauge theory either. There have been some attempts to describe general relativity as a gauge theory for diffeomorphisms, but this is impossible for several reasons. To begin with, general coordinate transformations take place on spacetime, not on the tangent space --- the fiber of the tangent bundle --- as it should be for a true gauge theory. In addition, general covariance by itself is empty of dynamical content in the sense that any relativistic equation, like for example Maxwell equation, can be written in a generally covariant form without any gravitational implication. There have also been some attempts to recast general relativity as a gauge theory for the Lorentz group. However, this is not possible either for different reasons. First, the spin connection of general relativity, as discussed in the previous section, is neither a true field variable nor a genuine gravitational connection. A second reason is that local Lorentz transformations are equivalent to general coordinate transformations in the sense that they give rise to the very same inertial connection. 

Indeed, observe that the inertial connection~(\ref{InerConn}), obtained by performing a {\em local Lorentz transformation}, and the inertial connection (\ref{InerConn2}), obtained by performing a {\em general coordinate transformation}, represent two different ways of expressing the very same inertial connection, as shown by Eq.~(\ref{STindIneCon2}).
Consciously or not, this equivalence is implicitly assumed in the metric formulation of general relativity. For example, it is a commonplace in many textbooks on gravitation to find the definition of a {\em locally inertial coordinate system}. Of course, the property of being or not inertial belongs to frames, not to coordinate systems. Such notion only makes sense if local Lorentz transformations and general coordinate transformations are considered on an equal footing. Then comes the point: since diffeomorphism is empty of dynamical meaning, and considering that it is equivalent to a local Lorentz transformation, the latter is also empty of dynamical meaning. One should not expect, therefore, any dynamical effect coming from a ``gaugefication'' of the Lorentz group.

On the other hand, there is a consistent rationale behind a gauge theory for the translation group. To begin with, remember that the source of gravitation is energy and momentum. From Noether's theorem, a fundamental piece of gauge theories \cite{kopo9}, we know that the energy-momentum tensor is conserved provided the source lagrangian is invariant under spacetime translations. If gravity is to be described by a gauge theory with energy-momentum as source, therefore, it must be a gauge theory for the translation group. This is similar to electrodynamics, whose source lagrangian is invariant under the one-dimensional unitary group $U(1)$, the gauge group of Maxwell theory.

\subsection{Gravity and the Quantum}
\label{sec:QuantGrav}
\index{Teleparallel gravity!and quantum gravity}

If general relativity is not a field theory in the usual sense of the term, the traditional approach of quantum field theory cannot be used in this case. In addition, due to the fact that general relativity is deeply rooted on the equivalence principle, its spin connection involves both gravitation and inertial effects. As a consequence, any approach to quantum gravity using this connection as field variable will necessarily include a quantization of the inertial forces --- whatever this may come to mean. Considering furthermore the divergent asymptotic behavior of inertial effects, like for example the centrifugal force, such approach is likely to face consistency problems. As a matter of fact, in the geometric approach of general relativity there is not a genuine gravitational variable to be quantized using the methods of quantum field theory. For these reasons, one should not expect to obtain a consistent quantum gravity theory from general relativity (Different arguments leading to the same conclusion can be found in Ref.~\cite{petkov}).

On the other hand, as a gauge theory for the translation group, teleparallel gravity is much more akin to a classical field theory than general relativity. It is, of course, different from the Yang-Mills type theories because of the soldering, which makes it a background-dependent field theory. In this theory, whereas inertial effects are represented by a Lorentz connection, the gravitational field is represented by a translational-valued connection, a legitimate gravitational variable in the usual sense of classical field theory. It is, for this reason, the variable to be quantized in any approach to quantum gravity. Taking into account that loop quantum gravity has a natural affinity with gauge theories \cite{lqg,Rovelli,Thiemann}, a quantization approach based on teleparallel gravity seems to be more consistent --- and of course much simpler due to the abelian character of translations. 

Still in connection to a prospective quantum theory for gravitation, it is important to remark that, differently from the geometrical approach of general relativity, the gauge approach of teleparallel gravity is not grounded on the equivalence principle \cite{wep}. In other words, it does not make use of the local equivalence between gravitation and inertial effects. As a consequence, it does not make use of ideal, local observers, as required by the strong equivalence principle, eliminating in this way the basic inconsistency with quantum mechanics, which presupposes real, dimensional observers \cite{vaxjo}. Of course, this is not enough to guarantee that a quantum version of teleparallel gravity will be a consistent theory, but can be considered an important conceptual advantage of teleparallel gravity.

\section{Final Remarks}

Although equivalent to general relativity, teleparallel gravity introduces new concepts into both classical and quantum gravity. For example, on account of the geometric description of general relativity, which makes use of the torsionless Levi-Civita connection, there is a widespread belief that gravity produces a curvature in spacetime. The universe as a whole, in consequence, should also be curved. However, the advent of teleparallel gravity breaks this paradigm: it becomes a matter of convention to describe the gravitational interaction in terms of curvature or in terms of torsion. This means that the attribution of curvature to spacetime is not an absolute, but a model-dependent statement. Notice furthermore that, according to teleparallel gravity, torsion has already been detected: it is responsible for all gravitational phenomena, including the physics of the solar system, which can be re-interpreted in terms of a force equation with torsion playing the role of force. A reappraisal of cosmology based on teleparallel gravity could provide a new way to look at the universe, eventually unveiling new perspectives not visible in the standard approach based on general relativity.

Not only cosmology, but many other gravitational phenomena would acquire a new perspective when analyzed from the teleparallel point of view. For instance, in teleparallel gravity there is a tensorial expression for the energy-momentum density of gravitation alone, to the exclusion of inertial effects. Gravitational waves would no longer be interpreted as the propagation of curvature-perturbation in the fabric of spacetime, but as the propagation of torsional field-strength waves. Furthermore, similarly to the teleparallel gauge potential, a fundamental spin-2 field should be interpreted, not as a symmetric second rank tensor, but as a translational-valued vector field \cite{spin2}. Most importantly, teleparallel gravity seems to be a much more appropriate theory to deal with the quantization of the gravitational field. We can then say that this theory is not just equivalent to general relativity, but a new way to look at all gravitational phenomena.

\section*{Acknowledgments}

The author would like to thank R. Aldrovandi for his long-standing collaboration in the development of the ideas presented in this monograph. He would like to thank also FAPESP, CAPES and CNPq for partial financial support.


\end{document}